\documentclass[12pt,tightenlines,eqsecnum,floats,showpacs,nofootinbib,amsmath,amssymb,aps,prd,superscriptaddress]{revtex4-1}

\usepackage{graphicx}
\usepackage{amsmath,amssymb}
\usepackage{hyperref}

\begin{document}
	
\title{Unique Fock quantization of a massive fermion field in a cosmological scenario}
	
\author{Jer\'onimo Cortez}
\email{jacq@ciencias.unam.mx}
\affiliation{Departamento de F\'isica, Facultad de Ciencias, Universidad Nacional Aut\'onoma de M\'exico, M\'exico D.F. 04510, M\'exico.}
\author{Beatriz Elizaga Navascu\'es}
\email{beatriz.elizaga@iem.cfmac.csic.es}
\affiliation{Instituto de Estructura de la Materia, IEM-CSIC, Serrano 121, 28006 Madrid, Spain.}
\author{Mercedes Mart\'in-Benito}
\email{m.martin@hef.ru.nl}
\affiliation{Institute for Mathematics, Astrophysics and Particle Physics, Radboud University Nijmegen,  Heyendaalseweg 135, NL-6525 AJ Nijmegen, The Netherlands.}
\author{Guillermo A. Mena Marug\'an} \email{mena@iem.cfmac.csic.es}
\affiliation{Instituto de Estructura de la Materia, IEM-CSIC, Serrano 121, 28006 Madrid, Spain.}
\author{Jos\'e M. Velhinho}
\email{jvelhi@ubi.pt}
\affiliation{Universidade da Beira Interior, Rua Marqu\^es d'\'Avila e Bolama, 6201-001, Covilh\~a, Portugal.}

\begin{abstract}

It is well known that the Fock quantization of field theories in general spacetimes suffers from an infinite ambiguity, owing to the inequivalent possibilities in the selection of a representation of the canonical commutation or anticommutation relations, but also owing to the freedom in the choice of variables to describe the field among all those related by linear time-dependent transformations, including the dependence through functions of the background. In this work we remove this ambiguity (up to unitary equivalence) in the case of a massive Dirac free field propagating in a spacetime with homogeneous and isotropic spatial sections of spherical topology. Two physically reasonable conditions are imposed in order to arrive to this result: (a) The invariance of the vacuum under the spatial isometries of the background, and (b) the unitary implementability of the dynamical evolution that dictates the Dirac equation. We characterize the Fock quantizations with a non trivial fermion dynamics that satisfy these two conditions. Then, we provide a complete proof of the unitary equivalence of the representations in this class under very mild requirements on the time variation of the background, once a criterion to discern between particles and antiparticles has been set.

\end{abstract}

	
\maketitle

\section{Introduction}
\label{sec:Intro}

Quantum field fheory (QFT) has proven to provide highly accurate and successful predictions within the standard model of particle physics, in absence of gravity. Indeed, the theoretical development of QFT in Minkowski spacetime has been crucial for explaining the phenomena observed in particle accelerators or astroparticle detectors. This formalism includes the assumption that the quantum theory incorporates the spacetime symmetries of the corresponding classical field theory, namely, the basic quantum structures are invariant under the transformations of the Poincar\'e group. However, flat spacetime is quite a specific background for the field theory, and in more general situations (allowed e.g. by general gelativity) no such a particular symmetry is present. Besides, and beyond any spacetime symmetry considerations, it is a well-known fact that there exists an infinite ambiguity when it comes to quantizing a linear field theory, ambiguity that is beyond the usual ones also present in traditional quantum mechanics (QM). In this latter case, and owing to the finiteness of the number of degrees of freedom, the joint criteria of strong continuity, unitarity, and irreducibility suffice to pick out a unique (up to unitary equivalence) quantum representation of the Weyl algebra on a separable Hilbert space, as stated by the Stone-von Neumann theorem \cite{svn}. However, field theories do possess an infinite number of degrees of freedom. Owing to this difference with standard QM, and even when restricting to free-field linear theories and to Fock-like representations of the field analogue of the Weyl algebra (or the corresponding field canonical commutation or anticommutation relations), one may construct an infinite number of inequivalent quantum theories \cite{wald}. This is a serious issue, inasmuch as those infinitely many possible quantum theories would provide different physical predictions. Remarkably, in the experimentally contrasted case of QFT in Minkowski spacetime, the mentioned symmetry requirements on the Fock representation of the canonical commutation or anticommutation relations turn out to select a \emph{unique} representation, thus removing this serious ambiguity \cite{birrell}.

For linear field theories, flat spacetime is not the only globally hyperbolic scenario for QFT where symmetry requirements have been put forward in order to select a privileged Fock quantization of the canonical commutation or anticommutation relations. For instance, in stationary (or static) spacetimes there is a well-established symmetry criterion to pick out a specific Fock representation \cite{wald,kay,ashmag}. Nonetheless, these results are not applicable if more general situations are investigated, where there might be fewer spacetime symmetries, and in particular not any time-translational symmetry at all. Specially relevant backgrounds of such type, from the physical point of view, are the nonstationary cosmological spacetimes. Luckily, recent investigations on the Fock quantization of real scalar fields propagating in such globally hyperbolic scenarios have proven that, in many of those situations, there still remain some physically meaningful criteria in order to select a unique class of unitarily equivalent representations. On the one hand, such criteria include the requirement of attaining a quantum theory that respects the \emph{spatial} isometries of the considered spacetime, with spatial sections that we take to have compact topology. On the other hand, the lack of any time-translation invariance is overcome with the requirement that the quantum dynamics be implementable as a unitary operator in the resulting Fock space (i.e., as a unitary endomorphism in that space), as it happens in traditional QM. This requirement allows us to use a Schr\"odinger picture in the quantum theory which is unitarily equivalent to the Heisenberg picture, commonly used in QFT. It therefore allows for a well-defined notion of ``particle creation'' on the vacuum over time, as it actually imposes ultraviolet regularity conditions \cite{unit}. These uniqueness results for the selection of a preferred Fock quantization of real free scalar fields in cosmological scenarios have been obtained in a variety of situations. The imposition of symmetry invariance and unitary implementability of the dynamics was first put forward in order to prove the uniqueness of the Fock quantization of gravitational waves in the inhomogeneous Gowdy cosmologies \cite{gowdy}. This proof has also been extended to the context of (test) scalar fields propagating in homogeneous and isotropic backgrounds with any kind of compact spatial sections, as long as the dimension of those sections is smaller than four \cite{compact}. A particularly interesting application of this result is the analysis of the Fock quantization of scalar (or tensor) cosmological perturbations around a flat Friedmann-Robertson-Walker (FRW) type of background \cite{torus}. In fact, that analysis reveals that, in such contexts, a unitary dynamics restricts the description of the fieldlike system by means of a very specific choice of field variable, obtained through a time-dependent scaling that includes part of the field evolution in the time dependence of the spacetime. Such a scaling is precisely the one employed in standard cosmology by the consideration of the Mukhanov-Sasaki variable \cite{Muk}. Moreover, the restriction applies as well to the field momentum, which is totally specified without any ambiguity in the possible addition to it of terms that are linear in the field configuration \cite{compact}.

Our aim now is to extend the uniqueness theorems proven so far for the Fock quantization of real scalar fields to the case of free fermion fields. The field of interest is thus the Dirac spinor. Such an object describes half-spin particles that are, together with gauge fields, the dominant type of matter in nature nowadays.  The interest in investigating the formalism underlying the description of fermion fields propagating in a curved spacetime goes beyond analyzing the possible role of fermions in cosmology. Indeed, it may prove to be useful in condensed matter systems such as graphene, where the low energy electronic excitations behave as if they were relativistic fermion fields \cite{graphene}.

As a first step towards removing the ambiguity in the Fock quantization of fermionic matter, in Ref. \cite{uf} we already considered the particular setting given by a massive Dirac spinor propagating in an FRW cosmology with closed spatial sections. This system was first studied in Ref. \cite{H-D}, where a particular Fock representation for the Dirac field, compatible with the symmetries of the field equations, was chosen. Concretely, the representation chosen in Ref. \cite{H-D} presents quite a specific structure with respect to the mode decomposition of the fermion field in terms of the eigenspinors of the Dirac operator on the spatial sections of the cosmological model, which are three-spheres ($S^{3}$). The main result of  Ref. \cite{uf} is the proof that this Fock quantization belongs to a family of equivalent representations that, being compatible with the symmetries of the field equations, also admits a unitary implementation of the quantum dynamics. This family of symmetry invariant representations is characterized by presenting a particular asymptotic behavior with respect to the eigenvalues of the Dirac operator on $S^{3}$. Such asymptotic behavior plays an important role in guaranteeing the unitary implementability of the dynamics \cite{uf}. Within this family, in Ref. \cite{uf} we took as the  \emph{reference} Fock representation the one obtained by considering the leading order in that asymptotic expansion with respect to the eigenvalues of the Dirac operator. Therefore, this quantization can be regarded as the simplest one that, while respecting the symmetries of the Dirac field equations under study, makes possible the unitary implementation of the fermion dynamics \cite{uf}.

In the present work, we generalize the previous results of Ref. \cite{uf} in two directions. First, we clarify the requirement of compatibility of the representation with the symmetries of the field equations used in Ref. \cite{uf}, showing that it suffices to impose invariance of the Fock quantization under the isometry group of $S^{3}$ in order to arrive at the desired mode structure for the representation. More importantly, we prove the uniqueness (up to unitary equivalence) of the representation under the imposed criteria, without further assumptions on the kind of asymptotic behavior. Indeed, we demonstrate that any Fock representation of the canonical anticommutation relations, invariant under the isometry group of $S^{3}$, that admits a nontrivial unitary implementation of the evolution, must be unitarily equivalent to our reference one, regardless of any additional restriction on its asymptotics (with respect to the eigenvalues of the Dirac operator) within the family of possible behaviors allowed by the unitarity of the dynamics. Such a unitary equivalence is established under very mild conditions on the time variation of the cosmological background, and once a convention on the notions of particle and antiparticle has been set. Remarkably, our analysis about the feasibility of a unitarily implementable and nontrivial evolution demonstrates as well the uniqueness of the time-dependent scaling of the fermion fields in the privileged definition of annihilation and creationlike variables selected by our criteria.

The structure of the paper is as follows. In Sec. \ref{sec:CS} we introduce the mathematical object that characterizes the Fock quantization of fermion fields and that then codifies the infinite ambiguity in its choice: The complex structure. We allow this structure to be explicitly time dependent, beyond the variation that is already due to its possible dynamical evolution. Afterwards, focusing the attention on the cosmological model under consideration, we will provide the general form of those complex structures that give rise to quantizations with a vacuum that is invariant under the spatial isometries of the considered spacetime. Sec. \ref{sec:Uniq} provides and analyzes the necessary and sufficient conditions that a complex structure must satisfy in order to admit a nontrivial unitary dynamics in the Fock space that it defines. With these conditions at hand, the most simple of those complex structures is selected as a reference. Then, we obtain a proof of the unitary equivalence of any other admissible complex structure. In this demonstration, the role of particles and antiparticles is discussed. Finally, in Sec. \ref{sec:Conclu} we summarize our results and comment on possible applications and extensions of them.

\section{Fermion complex structures}
\label{sec:CS}

A Fock representation of the canonical anticommutation relations of a fermion field is characterized by a complex structure. We introduce this object in the framework of a Dirac spinor $\Psi$ propagating in a general globally hyperbolic spacetime. The dynamics of the field is dictated by the first order Dirac equation in the corresponding spacetime. Specifically, we denote by ${S}=\{\Psi\}$ the complex linear space of solutions to that Dirac equation. Given the hyperbolicity of the background, each of those solutions is in a one-to-one correspondence with an initial value of the spinor field on a certain Cauchy surface \cite{Dimock}. Therefore, ${S}$ is isomorphic to the space of initial data for the Dirac equation. It is convenient to equip ${S}$ with the following natural, positive definite inner product \cite{Dimock}:
\begin{align}\label{inner}
(\Psi_1,\Psi_2)_S=\int d\tilde\mu\, \Psi^{+}_1 n^{\nu}e_{\nu}^{a}\gamma_{a} \Psi_2,
\end{align}
where the right-hand side is evaluated at a certain and arbitrary global time, and $d\tilde\mu$ is the integration measure on the spatial sections. Here, $\gamma_{a}$ (with $a=0,1,2,3$) are the constant Dirac matrices, that provide a representation of the Dirac-Clifford algebra
\begin{align}
\gamma_{a}\gamma_{b}+\gamma_{b}\gamma_{a}=2\eta_{ab}I,
\end{align}
where $I$ is the $4\times 4$ identity matrix and $\eta_{ab}$ is the Minkowski metric, taken with signature $\text{diag}\{-1,+1,+1,+1\}$. Also, we have used the notation $\Psi^{+}=\Psi^{\dagger}\gamma_{0}$ to denote the adjoint Dirac spinor, which satisfies the adjoint Dirac equation. Besides, $n^{\nu}$ are the spacetime components of the (unit, timelike, future-directed Lorentzian) normal to the spatial sections (with $\nu=0,1,2,3$), and $e^{a}_{\nu}$ denotes the tetrad. This inner product can be seen to be independent of the spatial section on which it is evaluated, so it is conserved under the evolution of the solutions of the Dirac equation \cite{Dimock}. In an analogous way, we may construct the space $\bar{S}$ as the complex conjugate of $S$, with the inner product given by the complex conjugate of \eqref{inner}. In the following, we denote complex conjugation with an overbar.

\subsection{General framework}
\label{subsec:General}

A complex structure $J: S\rightarrow S$ is defined as a real linear map with the property $J^2=-I$, and such that it leaves the inner product invariant \cite{Smatrix}. Any complex structure $J$ defines a splitting $S=S_J^+\oplus S_J^-$ of $S$ into two mutually complementary and orthogonal subspaces, $S^\pm_J=(S\mp iJS)/2$. In this way, $S_J^\pm$ are the eigenspaces of $J$ with eigenvalue $\pm i$. Analogously, we can define the complex structure $J$ as a linear map on $\bar{S}$. It then induces the splitting $\bar{S}=(\bar{S}_J)^+\oplus(\bar{S}_J)^-$, with $(\bar{S}_J)^\pm=(\bar{S}\mp iJ\bar{S})/2=\overline{(S_J^\mp)}$. With these decompositions at hand, the Fock quantization associated with $J$ is completely characterized by the assignation of the one-particle Hilbert space of particles to the completion of $S_J^+$ in the inner product \eqref{inner}, and the one-particle Hilbert space of antiparticles to the completion of $\overline{(S^-_J)}$. We denote these Hilbert spaces by $H_J^+$ and $\overline{H_J^-}$, respectively. The direct sum of these two Hilbert spaces gives rise to the one-particle Hilbert space $\mathcal H_J$ of the quantum theory, from which the antisymmetric Fock space is constructed. Therefore, different choices of complex structures provide different one-particle Hilbert spaces for the quantum theories that are in principle inequivalent to each other. Here resides the infinite ambiguity in the Fock representation of the canonical anticommutation relations of the Dirac field, ambiguity that we remove in this work.

In practice, the choice of a complex structure for the Fock quantization can be understood in terms of a specific choice of annihilation and creationlike variables for the classical field, to be quantized in terms of annihilation and creation operators. Indeed, the above decomposition is equivalent to picking out a set of particle annihilationlike variables, and a set of antiparticle creationlike ones, sets in terms of which the solutions of the Dirac equation are expressed. The subspace $S^{+}_J$ corresponds to the particle annihilationlike part of the solutions, whereas $S^{-}_J$ corresponds to the antiparticle creationlike part. Hence, we see that the choice of a complex structure is equivalent to that of a particular notion of specific particles and specific antiparticles, in the sense of determining their annihilation and creation operators (up to irrelevant linear transformations). Actually, the quantum theory can be uniquely specified by the corresponding vacuum state, defined as the normalized cyclic state which vanishes under the action of all the annihilation operators. The infinite ambiguity can then be equivalently traced back to the existing freedom in the choice of vacuum.

From a geometrical point of view, the domain $S$ of the complex structures can be described by local representatives, together with their complex conjugates, of cross sections of a spinor bundle on the considered spacetime. Specifically, given the principal bundle of Lorentzian frames on a globally hyperbolic spacetime \cite{Isham}, one may provide such a spacetime with a spin structure \cite{SGeom}. This procedure gives rise to an $SL(2,\mathbb{C})$-principal bundle, which may be regarded as the double covering of the bundle of Lorentzian frames, since $SL(2,\mathbb{C})$ is the double cover of the proper Lorentz subgroup. One may then associate a vector bundle to such an $SL(2,\mathbb{C})$-principal bundle, obtaining a spinor bundle \cite{SGeom}, which allows us to consistently describe half-spin particles via the local representatives of its cross sections, that are two-component objects. If the Weyl representation of the Dirac matrices is taken, each Dirac spinor is then a pair formed by one of those local representatives and by the complex conjugate of (the algebraic dual of) a different one. We call these local representatives ``two-component spinors''.

\subsection{Invariant complex structures}

The system considered in this work consists of a massive Dirac field minimally coupled to an expanding FRW cosmology, with spatial sections that have the topology of $S^{3}$. We call $\exp[{\alpha (\eta)}]$ the scale factor of this cosmology, where $\eta$ denotes conformal time. We work with the Weyl representation of the Dirac matrices so that, as pointed out above, each Dirac field is described in terms of two-component spinors $\phi^{A}$ and ${\bar{\chi}}_{A'}$. They describe the two parts of the Dirac field with well-defined and opposite chirality. We are using the index notation $A,B,...=0,1$ and $A',B',...=0',1'$ to denote the variables forming these two-component spinors, that we take to be of Grassmann type \cite{Berezin}. We follow the spinor conventions of Ref.  \cite{H-D}, summarized here in Appendix \ref{appendixA}.

Let us develop the framework needed to determine the general form that a complex structure must have to commute with the action of the isometry group of $S^{3}$ on the space of two-component spinors. Such complex structures naturally give rise to Fock representations that are invariant under this group. As a first step in this analysis, and in order to rigorously deal with the spatial dependence of the spinors, together with their properties under the action of the spatial isometry group, it is most convenient to perform a reduction of the structure group of the principal bundle of Lorentzian frames \cite{Isham}. This reduction is in practice realized by imposing a partial gauge fixing on the existing freedom in the choice of tetrads, such that all the attention is restricted to those that satisfy $e^{0}_{j}=0$, where $j=1,2,3$ denotes a spatial index on $S^{3}$. This gauge fixing condition is often known as time gauge, and it is commonly used in the study of fermion fields in globally hyperbolic spacetimes \cite{T-N}. After this partial gauge fixing has been imposed, the relevant objects for the bundle of frames are the triads, and this fact translates into a reduction of the structure group of such a bundle (the proper Lorentz subgroup) to $SO(3)$. Its double cover is then an $SU(2)$-principal bundle, to which the corresponding spinor bundle is associated. This spinor bundle turns out to describe, via the local representatives of its cross sections, two-component spinors over $S^{3}$, namely, over each of the spatial slices that foliate the considered cosmology. Once this reduction has been performed, the spinor fields may be expanded as follows \cite{H-D,uf}:
\begin{align}\label{harm1}
\phi_A(x)=&\frac{e^{-3\alpha(\eta)/2}}{2\pi}\sum_{npq}\breve{\alpha}_{n}^{pq}[m_{np}(\eta)\rho_{A}^{nq}(\textbf{x})+{\bar{r}}_{np}(\eta){\bar\sigma}_{A}^{nq}(\textbf{x})],
\\ \label{harm2}
{\bar\chi}_{A'}(x)=&\frac{e^{-3\alpha(\eta)/2}}{2\pi}\sum_{npq}\breve{\beta}_{n}^{pq}[{\bar{s}}_{np}(\eta){\bar\rho}_{A'}^{nq}(\textbf{x})+t_{np}(\eta)\sigma_{A'}^{nq}(\textbf{x})].
\end{align}
The sum over $p$ and $q$ is from 1 to $g_{n}=(n+1)(n+2)$, while $n$ is summed from 0 to infinity. On the other hand, $\rho_{A}^{np}(\textbf{x})$, ${\bar\sigma}_{A}^{np}(\textbf{x})$, and their complex conjugate fields are the spinor harmonics, that form a complete set for the expansion of any spinor field on $S^{3}$ \cite{H-D}. Analogous decompositions are performed for the complex conjugate fields $\bar{\phi}_{A'}$ and $\chi_{A}$. In all of them, the anticommuting nature of the spinors is captured by the Grassmann variables $m_{np}$, $\bar{r}_{np}$, $t_{np}$, $\bar{s}_{np}$ (and their complex conjugates). Furthermore, the constant coefficients $\breve{\alpha}_{n}^{pq}$ and $\breve{\beta}_{n}^{pq}$ are included for convenience, in order to avoid couplings between different values of $p$ when introducing the expansions in the Einstein-Dirac action. The specific values of these constants can be found in Appendix \ref{appendixA}.

The spinor harmonics are solutions to the following equations \cite{H-D}:
\begin{align}\label{dirac1}
-in_{AA'}\,e^{BA'j}\,{}^{(3)}\!D_{j}\rho^{np}_{B}=\frac{\omega_{n}}{2}\rho^{np}_{A}, \qquad
-in_{AA'}\,e^{BA'j}\,{}^{(3)}\!D_{j}{\bar{\sigma}}^{np}_{B}=-\frac{\omega_{n}}{2}{\bar{\sigma}}^{np}_{A},
\end{align}
\begin{align}\label{dirac2}
-in_{AA'}\,e^{AB'j}\,{}^{(3)}\!D_{j}{\bar{\rho}}^{np}_{B'}=-\frac{\omega_{n}}{2}{\bar{\rho}}^{np}_{A'},\qquad
-in_{AA'}\,e^{AB'j}\,{}^{(3)}\!D_{j}\sigma^{np}_{B'}=\frac{\omega_{n}}{2}\sigma^{np}_{A'},
\end{align}
where $\omega_{n}=n+3/2$, with $n\in\mathbb{N}$, and $p=1,...,g_{n}$. Here, ${}^{(3)}\!D_{j}$ is the local representative of the covariant derivative associated with the $SU(2)$ Levi-Civita connection on $S^{3}$, while $e^{AA'j}$ and $n^{AA'}$ denote, respectively, the spinor versions of the triad on $S^{3}$ and of the (unit, timelike, future-directed Lorentzian) normal to $S^{3}$. Let us notice that, owing to the partial gauge fixing, $n^{AA'}$ is simply minus $2^{-1/2}$ times the identity. Thus, Eqs. \eqref{dirac1} and \eqref{dirac2} are just the eigenvalue equations for the Dirac operator on $S^{3}$ with eigenvalues given by $\pm\omega_{n}$. The spinor harmonics then form the eigenspaces of this operator, each of which has dimension $g_{n}$ \cite{Yasushi}.

The isometry group of the three-sphere $S^{3}$ is $SO(4)$ or, equivalently, its double covering $\text{Spin}(4)=SU(2)\times SU(2)$, where the action on the points of $S^{3}$ is defined by Clifford multiplication \cite{Yasushi}. Specifically, $S^{3}$ is characterized as the homogeneous space such that the orbit of every one of its points under the action of $SO(4)$ covers the whole manifold. In other words, each point of $S^{3}$ is related to any other one via Clifford multiplication by an element of $\text{Spin}(4)$. An active $SO(4)$-transformation on the points of $S^{3}$, where the two-component spinors are evaluated, gives rise to a passive transformation on those spinors, via a representation of the isometry group $\text{Spin}(4)$ on the space of cross sections of the spinor bundle \cite{Yasushi}. Let us notice that, when considering the whole Dirac spinor, the action of this isometry group may be seen as two-block diagonal, with one block being the mentioned representation on $\phi^{A}$, and the other block being the corresponding complex conjugate representation on $\bar{\chi}_{A'}$. That action is unitary in the inner product \eqref{inner}, so each of these two blocks decomposes as the direct sum of irreducible representations of $\text{Spin}(4)$.

We are now ready to characterize those complex structures that commute with the action of $\text{Spin}(4)$ on the space of Dirac spinors. First of all, given the decompositions \eqref{harm1} and \eqref{harm2}, any complex structure can be regarded as an infinite-dimensional matrix in the basis formed by the modes $\{m_{np},\bar{r}_{np},t_{np}, \bar{s}_{np}\}$. Now, the properties of the eigenspaces of the Dirac operator on $S^{3}$ were thoroughly studied in Ref. \cite{Yasushi}. In particular, that work analyzes their relation with the irreducible decomposition of the isometry group when acting on the two-component spinors. Let us first restrict to the action of $\text{Spin}(4)$ on the space of two-component spinors with the quirality of $\phi^{A}$. There, via Frobenius reciprocity theorem \cite{Frob}, and as it was proven in Ref. \cite{Yasushi}, each of the eigenspaces of the Dirac operator on $S^{3}$ provides in fact each of the irreducible representations in which the action of $\text{Spin}(4)$ decomposes. Moreover, each of these irreducible representations appears with a multiplicity equal to one in the direct sum. This last fact, in particular, implies that the representation spaces spanned by
\begin{align}\label{eigensp}
\{\rho^{np}_{A}\}_{p=1,...,g_{n}} \qquad \text{and} \qquad \{{\bar{\sigma}}^{np}_{A}\}_{p=1,...,g_{n}},
\end{align}
for a given $n$, provide two inequivalent irreducible representations. Clearly, these representation spaces are equivalently spanned by the alternate basis obtained with the linear transformation $\breve{\alpha}_{n}$, used in the expansion \eqref{harm1}.  On the other hand, and following analogous arguments to those put forward in Ref. \cite{Yasushi}, the irreducible decomposition of the action of $\text{Spin}(4)$ on the space of spinors with the quirality of $\bar{\chi}_{A'}$ may be characterized in a similar way. Indeed, Frobenius reciprocity theorem guarantees that such action of $\text{Spin}(4)$ decomposes as a direct sum of irreducible representations that is componentwise equivalent to that in which its complex conjugate decomposed. More concretely, the irreducible representations provided by
\begin{align}
\{\bar{\rho}^{np}_{A'}\}_{p=1,...,g_{n}} \qquad \text{and} \qquad \{\sigma^{np}_{A'}\}_{p=1,...,g_{n}}
\end{align}
for each $n$ are respectively equivalent to those provided by \eqref{eigensp}. Again, these representation spaces are spanned as well by the basis reached with the transformation $\breve{\beta}_{n}$, introduced in Eq. \eqref{harm2}. Therefore, by applying Schur's lemma \cite{Reps} we can now state that any complex structure that commutes with the action of the isometry group of $S^{3}$ on the space of Dirac spinors, cannot mix modes $m_{np},\bar{r}_{np},t_{np}$, and $\bar{s}_{np}$ corresponding to different values of $n$. Furthermore, within the subspace corresponding to a given $n$, it cannot mix the modes $\{m_{np},\bar{s}_{np}\}$ with $\{t_{np},\bar{r}_{np}\}$, since, as we have seen, they provide inequivalent irreducible representations of $\text{Spin}(4)$. Let us consider, for a given value of $n$, the subspace spanned by the modes $\{m_{np},\bar{s}_{np}\}$. The restriction of the complex structure to such a subspace can then be characterized by four square maps between the two subspaces separately spanned by $\{m_{np}\}$ and $\{\bar{s}_{np}\}$. Since these two subspaces provide two irreducible representations of $\text{Spin}(4)$, Schur's lemma guarantees again that each of such maps must be proportional to the identity\footnote{To attain this result, notice that the constant linear transformations provided by $\breve{\alpha}_{n}$ and $\breve{\beta}_{n}$ actually guarantee that the two irreducible representations of $\text{Spin}(4)$ on the subspaces spanned by $\{m_{np}\}$ and $\{\bar{s}_{np}\}$ are the same. Such a property can be straightforwardly checked, e.g., by inspecting the Dirac equations for those modes, since they only relate modes $m_{np}$ and $\bar{s}_{np}$ with the same values of $n$ and $p$ \cite{uf}.}. A completely analogous argument is applied to the restriction of the complex structure to the subspace spanned by the modes $\{t_{np},\bar{r}_{np}\}$. Therefore, those restrictions can only relate modes that share the same value of the label $p$, and they cannot depend on that value.

In summary, we arrive at the following generic form for the complex structures that lead to invariant vacua of the Dirac field under the action of the isometry group corresponding to the spatial sections of the considered cosmology. In the basis provided by the tower of modes $\{m_{np},\bar{r}_{np},t_{np}, \bar{s}_{np}\}$, such complex structures can be regarded as block diagonal, with $2\times 2$ blocks that can at most mix the pairs of modes $(m_{np},\bar{s}_{np})$ or $(t_{np},\bar{r}_{np})$, for the same value of $n$ and $p$. Furthermore, the blocks that mix the pairs $(m_{np},\bar{s}_{np})$ are all equal for a given $n$, as it happens as well with those that mix $(t_{np},\bar{r}_{np})$. However, in principle, the blocks involving $(m_{np},\bar{s}_{np})$ need not be equal to those involving $(t_{np},\bar{r}_{np})$. From now on, every complex structure with these properties is called \emph{invariant} complex structure.

\section{Uniqueness of the quantization}
\label{sec:Uniq}

In the remaining discussion,we restrict our attention to Fock quantizations of the Dirac field that are determined by invariant complex structures. Within such a subset of quantizations, Ref. \cite{uf} selected a particularly simple one that displays a physically appealing property: It admits an implementation of the fermion dynamics as a unitary operator on the Fock space of the quantum theory. However, this reference quantization is not the only invariant one that allows for such a unitarily implementable evolution. There exist infinitely many invariant vacua that satisfy these properties. Indeed, in Ref. \cite{uf} we already proved that the reference quantization selected in that work belongs to a class of unitarily equivalent Fock quantizations which, apart from being invariant under the group of spatial isometries and allowing for a unitary implementation of the dynamics, are characterized by having complex structures that admit a specific asymptotic behavior with respect to the eigenvalues of the Dirac operator on $S^3$. However we did not demonstrate the uniqueness of the quantization, in the sense that we did not prove that any other complex structure compatible with the criteria of invariance under the isometry group and of unitary implementability of the dynamics is necessarily equivalent to the class of complex structures studied in Ref. \cite{uf}. We still have the possibility of finding inequivalent quantizations among those that do not possess the asymptotics of that class.  At this point, then, our aim is to elucidate whether all the Fock representations, selected by the symmetry and unitary dynamics requirements, are indeed unitarily equivalent to our reference one. If this is the case, then the physics that they describe, at the level of the canonical anticommutation relations, will all be the same, and the infinite ambiguity in the quantization will be removed.

\subsection{Unitary dynamics}
\label{refe}

Given a specific invariant complex structure, we  call respectively $a^{(x,y)}_{np}$ and $b^{(x,y)}_{np}$ the particle and antiparticle annihilationlike variables that it selects. The creationlike variables $a_{np}^{(x,y)\dagger}=\bar{a}_{np}^{(x,y)}$ and $b_{np}^{(x,y)\dagger}=\bar{b}_{np}^{(x,y)}$ are their complex conjugates. {If we use the notation $\{x_{np},y_{np}\}$ to describe either of the \emph{ordered} sets $\{m_{np},s_{np}\}$ or $\{t_{np},r_{np}\}$, then, from the discussion in Sec. \ref{sec:CS}, it is clear that any invariant complex structure leads, at any time $\eta$, to annihilation and creationlike variables of the form:
\begin{align}\label{anni}
\begin{pmatrix} a_{np}^{(x,y)} \\ b^{(x,y)\dagger}_{np} \\ a^{(x,y)\dagger}_{np} \\ b_{np}^{(x,y)}\end{pmatrix}_{\!\!\eta}=\begin{pmatrix}f_{1}^ {n}(\eta) & f_{2}^{n}(\eta) & 0 & 0 \\ g_{1}^ {n}(\eta) & g_{2}^{n}(\eta) & 0 & 0 \\ 0 & 0 & \bar{f}^{n}_{1}(\eta) & \bar{f}^{n}_{2}(\eta) \\ 0 & 0 & \bar{g}^{n}_{1}(\eta) & \bar{g}^{n}_{2}(\eta)\end{pmatrix}\begin{pmatrix} x_{np} \\ \bar{y}_{np} \\ \bar{x}_{np} \\ y_{np}\end{pmatrix}_{\!\!\eta}.
\end{align}
Since we allow for complex structures that may depend explicitly on time, we have contemplated the possible time variation of the linear combinations that define the annihilation and creationlike variables. Consequently, the coefficients $f_{l}^{n}$ and $g_{l}^{n}$ (with $l=1,2$) are, in principle, time functions, as our notation indicates. Besides, we have used a subindex $\eta$ in column vectors to denote evaluation at that value of the conformal time.

Although not explicitly displayed here, in order not to complicate the notation, the time-dependent coefficients $f_{l}^{n}$ and $g_{l}^{n}$, as well as their complex conjugates $\bar{f}^{n}_{l}$ and $\bar{g}^{n}_{l}$, may in general differ for the mode pairs $(m_{np},\bar{s}_{np})$ and $(t_{np},\bar{r}_{np})$. However, in both cases they must obey the relations
\begin{align}\label{sympl}
|f_{1}^{n}|^{2}+|f_{2}^{n}|^{2}=1, \qquad |g_{1}^{n}|^{2}+|g_{2}^{n}|^{2}=1, \qquad f_{1}^{n}\bar{g}^{n}_{1}+f_{2}^{n}\bar{g}^{n}_{2}=0,
\end{align}
so that the considered variables are indeed annihilation and creationlike, from the Hamiltonian point of view \cite{H-D,uf}. In particular, these relations imply that
\begin{align}\label{fgrel}
g_{1}^{n}=\bar{f}_{2}^{n}e^{iG^{n}}, \qquad g_{2}^{n}=-\bar{f}^{n}_{1}e^{iG^{n}},
\end{align}
with $G^{n}$ being a certain phase, possibly time dependent.

It can be seen that the Dirac equations for the fermion modes give rise to the following Bogoliubov transformation that implements the dynamics on the annihilation and creationlike variables, from a given initial time $\eta_{0}$ to any other time $\eta$ \cite{uf}:
\begin{align}\label{bog}
\begin{pmatrix} a_{np}^{(x,y)} \\ b^{(x,y)\dagger}_{np} \\ a^{(x,y)\dagger}_{np} \\ b_{np}^{(x,y)}\end{pmatrix}_{\!\!\eta}=B_{n}(\eta,\eta_{0})\begin{pmatrix} a_{np}^{(x,y)} \\ b^{(x,y)\dagger}_{np} \\ a^{(x,y)\dagger}_{np} \\ b_{np}^{(x,y)}\end{pmatrix}_{\!\!\eta_{0}},
\end{align}
with the block-diagonal form
\begin{align}
B_{n}=\begin{pmatrix} \mathcal{B}_{n} & 0 \\ 0 & \mathcal{\bar{B}}_{n} \end{pmatrix}, \qquad \mathcal{B}_{n}=\begin{pmatrix} \alpha_{n}^{f} & \beta_{n}^{f} \\ \beta_{n}^{g} & \alpha_{n}^{g} \end{pmatrix},
\end{align}
where $\alpha_{n}^{f},\alpha_{n}^{g}$, $\beta_{n}^{f}$, and $\beta_{n}^{g}$ are coefficients (dependent  on $\eta$ and $\eta_0$) with an expression that can be found in Ref. \cite{uf}. For future reference, we give here the explicit formula for the beta coefficients:
\begin{align}\label{beta1}
\beta_{n}^{h}(\eta,\eta_0)=&\frac{1}{h_{1}^{n,0}k_{2}^{n,0}-h_{2}^{n,0}k_{1}^{n,0}}\Bigg\{\left[-h_{1}^{n}\bigg(h_{2}^{n,0}+\Gamma_n h_{1}^{n,0}\bigg)e^{i\int \Lambda^{1}_{n}}+\bar\Gamma_n h^{n}_{2}h_{2}^{n,0}e^{\Delta\alpha} e^{i\int\bar{\Lambda}^{2}_{n}}\right] e^{i\omega_{n}\Delta\eta}\nonumber\\& +\left[h_{2}^{n}\bigg(h_{1}^{n,0} -\bar\Gamma_n h_{2}^{n,0}\bigg)e^{-i\int \bar{\Lambda}^{1}_{n}}+\Gamma_n h^{n}_{1}h_{1}^{n,0}e^{\Delta\alpha}e^{-i\int\Lambda^{2}_{n}}\right] e^{-i\omega_{n}\Delta\eta}\Bigg\}.\end{align}
Here, $\{h,k\}= \{f,g\}$ as a set, and we are using the notation $\Delta\eta=\eta-\eta_{0}$ and $\Delta\alpha=\alpha-\alpha_{0}$, with $\alpha_{0}=\alpha(\eta_{0})$, and
\begin{align}\label{gamma}
\Gamma_{n}=\frac{me^{\alpha_{0}}}{2\omega_{n}+i\alpha_{0}'},
\end{align}
where the prime stands for derivation with respect to the conformal time. The integrals in the above expression are in conformal time, in the interval $[\eta_{0},\eta]$, and $\Lambda^{j}_{n}$ are some time-dependent functions which are of order $\mathcal{O}(\omega_{n}^{-1})$ in the asymptotic limit of large values of $\omega_{n}$, as long as $\alpha$ and its derivatives (up to third order) exist and are integrable in every closed interval $[\eta_{0},\eta]$ \cite{uf}.  Besides, to simplify the notation, we have omitted the dependence of the functions on $\eta$, and distinguished evaluation at $\eta_{0}$ with the superscript $0$ (preceded by a comma).

Let us now provide the necessary and sufficient conditions that any complex structure must satisfy in order to have a unitarily implementable dynamics in the quantum theory that it defines. It is well known that any linear canonical transformation of a fieldlike system is implementable as a unitary operator in the Fock space if and only if its antilinear part is of the Hilbert-Schmidt class \cite{Shale}. For the dynamical Bogoliubov transformation defined by the sequence of matrices $B_{n}(\eta,\eta_{0})$, such a condition is satisfied if and only if the sums
\begin{align}\label{ucond}
\sum_{n}g_{n}|\beta_{n}^{f}(\eta,\eta_0)|^{2} \qquad \text{and} \qquad \sum_{n}g_{n}|\beta_{n}^{g}(\eta,\eta_0)|^{2}
\end{align}
are convergent for all times $\eta$ in the evolution. Recall that the degeneracy of the eigenvalues $\pm \omega_{n}$ of the Dirac operator is $g_{n}=(n+1)(n+2)=\omega_{n}^{2}-1/4$. It is therefore clear that conditions \eqref{ucond} imply in particular that, in the region of large $\omega_{n}$, the beta coefficients must be negligible e.g. compared to $\omega_{n}^{-1}$ for all times $\eta$. This necessary condition for a Fock quantization to admit a unitary dynamics turns out to translate into a very specific behavior of the coefficients $f_{l}^{n}$ and $g_{l}^{n}$. Such a behavior affects both their dependence on the eigenvalues of the Dirac operator, as well as the time-dependent scaling that they introduce in the different parts of the fermion field. In order to show this fact, let us denote $\{l,\tilde{l}\}= \{1,2\}$ as a set. Then, taking into account relations \eqref{sympl} and \eqref{fgrel}, the following scenarios can be distinguished (at least in a sufficiently short time interval beyond $\eta_0$), in principle not necessarily mutually exclusive:

\begin{itemize}
\item[i)] If $\omega_{n}^{-1}$ is negligible compared to $h_{l}^{n}$ in the asymptotic limit of large $\omega_{n}$ for an infinite subset of the natural numbers, $n\in\mathbb{N}_{l}^{\uparrow}$. Since
\begin{align}
h_{\tilde{l}}^{n}=e^{iH^{n}_{\tilde{l}}}\sqrt{1-|h_{l}^{n}|^{2}},
\end{align}
with $H^{n}_{\tilde{l}}$ being some possibly time-dependent phase, we see that $h_{\tilde{l}}^{n}$ is of the same order as $h_{l}^{n}$ or either of unit order, whatever is the largest. The only possible exception happens perhaps if $h_{l}^{n}$ is of the order of the unity, when  $h_{\tilde{l}}^{n}$ may be smaller; we comment on this exceptional case below.

On the other hand, it is easy to check that  the denominator in Eq. \eqref{beta1} has unit norm \cite{uf}. Besides, we notice that $\Gamma_n$ is of the order of $\omega_n^{-1}$, and we recall that $\beta^{h}_{n}(\eta,\eta_{0})$ has to be negligible with respect to $\omega_{n}^{-1}$ for all times $\eta$, because the dynamics is assumed to be unitarily implementable. Taking into account all these behaviors, it follows from Eq. \eqref{beta1} that
\begin{align}\label{hdynequal}
\frac{h_{l}^{n}}{h_{l}^{n,0}}=\frac{h_{\tilde{l}}^{n}}{h_{\tilde{l}}^{n,0}}e^{i(-1)^{l}2\omega_{n}\Delta\eta},
\end{align}
up to terms negligible compared to $\omega_{n}^{-1}$, for all $\eta$ and all $n\in\mathbb{N}_{l}^{\uparrow}$ bigger than a certain $n_{l}\geq 0$. However, this is not possible unless the time dependence of the phases $H_{l}^{n}$ and $H_{\tilde{l}}^{n}$, of $h_{l}^{n}$ and $h_{\tilde{l}}^{n}$ respectively, are fixed so as to absorb the dominant dynamical contribution in the phases of the fermion modes, hence trivializing the dynamics and rendering it equal to the identity evolution at the considered order. In more detail, condition \eqref{hdynequal} requires that $H^{n}_{l}-H^{n,0}_{l}=H_{\tilde{l}}^{n}-H_{\tilde{l}}^{n,0} +(-1)^{l}2\omega_{n}\Delta\eta +2\pi k_n$, with $k_n\in \mathbb{Z}$. We rule out this possibility (that leads to a trivial dynamics), e.g. by demanding that the time dependence of the dominant contribution in our coefficient be mode independent.

\item[ii)] If $h_{l}^{n}$ is negligible compared to $\omega^{-1}_{n}$ in the asymptotic limit of large $\omega_{n}$ for an infinite subset of the natural numbers, $n\in\mathbb{N}_{l}^{\downarrow}$. In this case, one can check from Eq. \eqref{beta1} that a unitarily implementable dynamics can never be reached, leading to a contradiction, because the assumption that $\beta_{n}^{h}(\eta,\eta_{0})$ be negligible versus $\omega_{n}^{-1}$ implies the impossible relation
\begin{align}\label{impocondi}
e^{\Delta\alpha+i(-1)^{l+1} 2 \omega_{n}\Delta\eta}=
1,
\end{align}
up to negligible terms,	for all times $\eta$ and for all $n\in\mathbb{N}_{l}^{\downarrow}$ bigger than a certain $n_{l}\geq 0$. We have used the mentioned fact that the functions $\Lambda_n^{j}$ are $\mathcal{O}(\omega_n^{-1})$.

\item[iii)] If $h_{l}^{n}$ is of order  $\omega_{n}^{-1}$ in the regime of large $\omega_{n}$ for an infinite subset of the natural numbers, $n\in\mathbb{N}_{l}$. We then have $h_{l}^{n}\sim p^{n}+o(\omega_{n}^{-1})$, where $p^{n}=\mathcal{O}(\omega_n^{-1})$ [and $o(\omega_{n}^{-1})$ denotes terms negligible compared to $\omega_{n}^{-1}$]. Employing that $\Gamma_n=m\exp{(\alpha_0)} /(2 \omega_n)$ at dominant order, and following similar arguments as in the two other cases discussed above, it is easy to conclude from Eq. \eqref{beta1} that, for all times $\eta$ and all $n\in\mathbb{N}_{l}$ bigger than a certain $n_{l}\geq 0$, we must have
\begin{align}
p^{n}=(-1)^{l+1}\frac{me^{\alpha}}{2\omega_{n}}e^{iH_{\tilde{l}}^{n}}+\left[p^{n,0}+(-1)^{l}\frac{me^{\alpha_{0}}}{2\omega_{n}}e^{iH_{\tilde{l}}^{n,0}}\right]
e^{i(-1)^{l}2\omega_{n}\Delta\eta+iH_{\tilde{l}}^{n}-iH_{\tilde{l}}^{n,0}},
\end{align}
where $H_{\tilde{l}}^{n}$ is, again, the phase of $h_{\tilde{l}}^{n}$. Therefore, if we discard the possibility of absorbing the dominant time variation of the phase of the fermions  in a term of $p^{n}$, asking instead, e.g., that the time dependence of the norm of the leading contribution to $h_{l}^n$ factorizes, the coefficient $h_{l}^{n}$ must necessarily behave as
\begin{align}\label{unith}
h^{n}_{l}=(-1)^{l+1}\frac{me^{\alpha}}{2\omega_n}e^{iH^{n}_{\tilde{l}}}+o(\omega_{n}^{-1}),
\end{align}
for all $n\in\mathbb{N}_{l}$, in the asymptotic limit of large $\omega_{n}$.

\item[iv)] Finally, let us comment on the case in which $h_{l}^{n}$ is of unit order and $h_{\tilde{l}}^{n}$ is of smaller order, let us say for a subset of natural numbers $n\in\mathbb{N}_{\tilde{l}}$. In this situation, it is convenient to reverse the roles of $h_{l}^{n}$ and $h_{\tilde{l}}^{n}$ and repeat the analysis above, this time applied to $h_{\tilde{l}}^{n}$ (and with the first scenario  restricted to the subcase in which $h_{\tilde{l}}^{n}$ is already negligible compared to the unity). Then, one concludes that the only acceptable possibility is that $h_{\tilde{l}}^{n}$ displays a behavior like that in \eqref{unith} (with $l$ replaced  with $\tilde{l}$) for $n\in\mathbb{N}_{\tilde{l}}$, if this set of numbers is unbounded.
\end{itemize}

Let us now discuss the behavior allowed by Eq. \eqref{unith} in more detail. We call $\vartheta^{n}_{h,l}$ the term in that equation that is negligible compared to $\omega_{n}^{-1}$. A simple inspection of the dynamical beta coefficients \eqref{beta1} and the conditions \eqref{ucond}, taking into account the asymptotic behavior that we have already commented on for the different functions that appear in our expressions, allows us to deduce that a nontrivial unitarily implementable dynamics in the considered quantization with invariant complex structure is accomplished \emph{if and only if} the sequence
\begin{align}\label{sqssequence}
\vartheta^{n,0}_{h,l}-\vartheta^{n}_{h,l}e^{i(-1)^{l+1}2\omega_{n}\Delta\eta -iH_{\tilde{l}}^{n}+iH_{\tilde{l}}^{n,0}},
\end{align}
with $n\in\mathbb{N}_{l}$, is square summable at all studied times, including degeneracy. If the sequence formed by the terms $\vartheta^{n,0}_{h,l}$ at the initial time is square summable, this is equivalent to demanding that the corresponding sequence formed by $\vartheta^{n}_{h,l}$ (with $n\in\mathbb{N}_{l}$ ) be square summable at all the instants of time under consideration. Otherwise, if the sequence given by $\vartheta^{n,0}_{h,l}$ has a part $\tilde{\vartheta}^{n,0}_{h,l}$ that is not square summable, one would need to compensate it at every time with the time-dependent factor in Eq. \eqref{sqssequence}, something that completely fixes the possible time variation of the contribution in $\vartheta^{n}_{h,l}$ that is not square summable. If we call this contribution $\tilde{\vartheta}^{n,0}_{h,l}\exp{(i\theta^{n}_{h,l})}$, we should have $\theta^{n}_{h,l}=H_{\tilde{l}}^{n}-H_{\tilde{l}}^{n,0}+(-1)^{l} 2\omega_{n}\Delta\eta +2\pi k_n$, with $k_n\in\mathbb{Z}$. Therefore, the dominant phase in the dynamics of the fermion modes would be absorbed in the time dependence of the complex structure (via $\theta^{n}_{h,l}$ and $H_{\tilde{l}}^{n}$), trivializing the evolution at the level of the terms that are not square summable. If we discard this possibility, e.g. by requiring that the time dependence of the discussed contributions be given only by mode-independent functions, we conclude that we must necessarily have that  $\vartheta^{n}_{h,l}$ is square summable, with the degeneracy included, over the subset ${n\in\mathbb{N}_{l}}$.

Finally, let us notice that, once $h_{l}^{n}$ is fixed as either $f_{l}^{n}$ or $g_{l}^{n}$, then, the other beta functions, which can be called $\beta^{k}_{n}(\eta,\eta_{0})$ with our notation, can be checked to coincide, in complex norm, with $\beta^{h}_{n}(\eta,\eta_{0})$, as we showed in Ref. \cite{uf}. Therefore, if the sequence $\beta^{h}_{n}$ is square summable (with the known degeneracy), so is $\beta^{k}_{n}$.

In summary, the necessary and sufficient conditions for a complex structure to define a Fock quantization that is symmetry invariant and has a unitarily implementable dynamics are that the corresponding functions $h_{l}^{n}$ be asymptotically of the form \eqref{unith}, and that the terms of order $o(\omega_n^{-1})$ form a sequence $\{\vartheta^{n}_{h,l}\}_{n\in\mathbb{N}_{l}}$ that is square summable, including degeneracy. Here, we have split the natural numbers, up to a finite collection of them, into two infinite subsets $\mathbb{N}_{1}$ and $\mathbb{N}_{2}$, such that $\mathbb{N}=\mathbb{N}_{1}\cup\mathbb{N}_{2}$, allowing for the possibility that one of these subsets be empty. Our statement is true as long as the different time-dependent phases in the definition of the annihilation and creationlike variables that contribute to terms that are not square summable are not fixed so as to trivialize the dominant dynamical contribution in the phases of the fermion modes, with respect to an asymptotic expansion in terms of $\omega_{n}$. Actually, a subclass of complex structures of the kind which we have found that satisfy the symmetry and unitary evolution criteria was the one contemplated in Ref. \cite{uf}, characterized by the additional requirement that the next-to-leading order in $h_{l}^{n}$ be $\mathcal{O}(\omega_{n}^{-2})$. In this sense, the analysis performed here generalizes this subclass and uniquely characterizes both the dominant asymptotic behavior and its time-dependent scaling for any invariant Fock quantization of the Dirac field that admits a nontrivial and unitarily implementable fermion dynamics.

\subsection{Uniqueness}
\label{proof}

We now prove that there exists indeed only one equivalence class of complex structures uniquely selected by the criteria of invariance under the spatial isometries of $S^3$, and of unitary implementability of a (nontrivial) quantum dynamics. We show that this result is valid once the convention for particles and antiparticles has been settled.

In light of the results obtained in the previous section, it suffices to restrict our attention to invariant complex structures that lead to Fock quantizations which allow for a unitary dynamics, without trivializing it. Specifically, let us restrict to sets of annihilation and creationlike variables of the type \eqref{anni} with $h_{l}^{n}$ being a function of the asymptotic form \eqref{unith}, and such that the corresponding sequences formed by $\vartheta^{n}_{h,l}$ (for
$n\in\mathbb{N}_{l}$) are square summable. Within this class of complex structures, we fix a \emph{reference} structure $J_{R}$: The one that selects annihilation and creationlike variables with
\begin{align}\label{reference}
f_{1}^{n}=\frac{me^{\alpha}}{2\omega_{n}},\qquad f_{2}^{n}=\sqrt{1-(f_{1}^{n})^2},\qquad g_{1}^{n}=f_{2}^{n}, \qquad g_{2}^{n}=-f_{1}^{n},
\end{align}
both for the pairs $(m_{np},\bar{s}_{np})$ and for $(t_{np},\bar{r}_{np})$. This choice is obviously the simplest one that satisfies the necessary and sufficient conditions for admitting a nontrivial unitary quantum dynamics.

Let us call $\tilde{J}$ any other complex structure that defines a Fock quantization with a nontrivial and unitarily implementable dynamics. A complex structure of this kind is characterized by some annihilation and creationlike variables $\{\tilde{a}_{np}^{(x,y)},\tilde{b}_{np}^{(x,y)},\tilde{a}_{np}^{(x,y)\dagger},\tilde{b}_{np}^{(x,y)\dagger}\}$ with coefficients $\tilde{f}_{l}^{n}$ and $\tilde{g}_{l}^{n}$, as those given in Eqs. \eqref{anni}-\eqref{fgrel}, such that either $\tilde{f}_{l}^{n}$ or $\tilde{g}_{l}^{n}$ are of the asymptotic form \eqref{unith}, and such that the corresponding sequences formed by $\vartheta^{n}_{\tilde{h},l}$ are square summable. As we showed in Ref. \cite{uf}, the relation between $\tilde{J}$ and our reference complex structure $J_{R}$ at a certain time $\eta$ is given by
\begin{align}\label{bogK}
\begin{pmatrix} \tilde a_{np}^{(x,y)} \\ \tilde b^{(x,y)\dagger}_{np} \\ \tilde a^{(x,y)\dagger}_{np} \\ \tilde b_{np}^{(x,y)}\end{pmatrix}_{\!\!\eta}=K_{n}(\eta)\begin{pmatrix} a_{np}^{(x,y)} \\ b^{(x,y)\dagger}_{np} \\ a^{(x,y)\dagger}_{np} \\ b_{np}^{(x,y)}\end{pmatrix}_{\!\!\eta},
\end{align}
with\footnote{Although not specified with our notation, the entries in the matrix $K_{n}$ might be different for the annihilation variables $(\tilde a^{(m,s)}_{np},\tilde b^{(m,s)}_{np})$ and $(\tilde a^{(t,r)}_{np},\tilde b^{(t,r)}_{np})$. We avoid this distinction in the subsequent analysis, since both cases are dealt with in exactly the same way.}
\begin{align}\label{Kn}
K_{n}=\begin{pmatrix} \mathcal{K}_{n} & 0 \\ 0 & \mathcal{\bar{K}}_{n} \end{pmatrix}, \qquad \mathcal{K}_{n}=\begin{pmatrix} \kappa_{n}^{f} & \lambda_{n}^{f} \\ \lambda_{n}^{g} & \kappa_{n}^{g} \end{pmatrix},
\end{align}
and
\begin{align}\label{kl}
\kappa^{h}_{n}=\frac{\tilde{h}^{n}_{2}k_{1}^{n}-
\tilde{h}^{n}_{1}k_{2}^{n}}{h_{2}^{n}k_{1}^{n}-h_{1}^{n}k_{2}^{n}},\qquad
\lambda^{h}_{n}=\frac{\tilde{h}_{1}^{n}h_{2}^{n}-\tilde{h}_{2}^{n}h_{1}^{n}}{h_{2}^{n}k_{1}^{n}-h_{1}^{n}k_{2}^{n}},
\end{align}
where we recall that $\{h,k\}= \{f,g\}$ as a set. Now, the Fock representations determined by $\tilde J$ and by $J_R$ are unitarily equivalent if and only if the time-dependent transformation determined by the sequence of matrices $K_{n}$ can be implemented as a unitary operator in the quantum theory, a fact that in turn is true if and only if its antilinear part is a Hilbert-Schmidt operator \cite{Shale}, namely when
\begin{align}\label{ucond2}
\sum_{n}g_{n}|\lambda_{n}^{f}(\eta)|^{2}<\infty \qquad \text{and} \qquad \sum_{n}g_{n}|\lambda_{n}^{g}(\eta)|^{2}<\infty,
\end{align}
for all times $\eta$. We show in what follows that any of the invariant complex structures $\tilde{J}$, that defines a quantum theory for which a nontrivial unitary dynamics can be implemented, is indeed unitarily equivalent to our reference one, once a convention of particles and antiparticles has been set, thus proving the uniqueness of the quantization. Two possible cases can be distinguished in our analysis, given the two {existing possibilities that either the coefficients $\tilde{f}_{l}^{n}$ or the coefficients $\tilde{g}_{l}^{n}$ adopt the form \eqref{unith}. We analyze only one of them, because the other case can be dealt with in a completely parallel manner, owing to relations \eqref{fgrel}.

Let us thus admit that the coefficients $\tilde{f}_{1}^{n}$ and $\tilde{f}_{2}^{n}$ are, respectively, of the asymptotic form \eqref{unith} for $n\in\mathbb{N}_{1}$ and for $n\in\mathbb{N}_{2}$. Recall that the union of these two subsets is the entire set of the natural numbers (apart maybe from a finite subset, irrelevant for our considerations), and that one of these subsets may be void. Then, from  definition \eqref{kl}, one can check that, for $n\in\mathbb{N}_{1}$ and in the asymptotic limit of large $\omega_{n}$,
\begin{align}\label{lambdafg}
\lambda^{f}_{n}=\vartheta^{n}_{\tilde{f},1}+\mathcal{O}(\omega_{n}^{-2}), \qquad \lambda^{g}_{n}=-\bar{\lambda}^{f}_{n}e^{i\tilde{G}^{n}}=-\bar{\vartheta}^{n}_{\tilde{f},1}
e^{i\tilde{G}^{n}}+\mathcal{O}(\omega_{n}^{-2}),
\end{align}
with $\tilde{G}^{n}$ being a certain phase. Since the sequence formed by $\vartheta^{n}_{\tilde{f},1}$ for $n\in\mathbb{N}_{1}$ is square summable with our hypotheses (taking into account the degeneracy $g_n=\omega_n^2-1/4$), then, if $\mathbb{N}_{2}$ is void or has a finite number of elements, we see that conditions \eqref{ucond2} are immediately satisfied, and the quantizations defined by $\tilde{J}$ and $J_{R}$ are unitarily equivalent. Consider the case in which $\mathbb{N}_{2}$ is not bounded from above, so that its cardinal is infinite. It is easy to see that both $\lambda_{n}^{f}$ and $\lambda_{n}^{g}$ are $\mathcal{O}(1)$, for $n\in\mathbb{N}_{2}$ in the asymptotic limit of large $\omega_{n}$. Therefore, the sequences that they provide are not square summable, since the degeneracy $g_{n}$ increases as $\omega_n^2$. In this sense, the quantizations defined by $\tilde{J}$ and $J_{R}$ would not be unitarily equivalent. However, such inequivalence can be understood as artificially arising from the fact that the complex structure $\tilde{J}$ defines a convention for the concept of particles and antiparticles which is completely reversed, for an infinite number of modes, in comparison with the convention corresponding to $J_{R}$ \cite{uf}. Once both conventions are reconciled, the physical phenomena described by both quantizations would be exactly the same. Indeed, suppose that we had taken as the reference complex structure not the one introduced above, $J_R$, but another one, that we call $\tilde J_R$, obtained from $J_R$ by switching the convention of particles and antiparticles, discussed in Sec.\ref{subsec:General}, for all modes corresponding to $\mathbb{N}_{2}$ from some $n_{2}\geq 0$ onwards. Such a change in the convention amounts in practice to the interchange $f_{l}^{n}\leftrightarrow g_{l}^{n}$ \cite{uf}. The lambda coefficients relating $\tilde J_R$ and $\tilde{J}$ would then be given asymptotically by Eq. \eqref{lambdafg} for $n\in\mathbb{N}_{1}$, whereas
\begin{align}
\tilde{\lambda}_{n}^{f}=\vartheta^{n}_{\tilde{f},2}+\mathcal{O}(\omega_{n}^{-2}),\qquad \tilde{\lambda}_{n}^{g}=\bar{\tilde{\lambda}}^{f}_{n}e^{i\tilde{G}^{n}}=\bar\vartheta^{n}_{\tilde{f},2}
e^{i\tilde{G}^{n}}+\mathcal{O}(\omega_{n}^{-2}),
\end{align}
for $n\in\mathbb{N}_{2}$. But now, according to our hypotheses,  $\vartheta^{n}_{\tilde{f},2}$ is square summable, including degeneracy, over $n\in\mathbb{N}_{2}$. Hence, conditions \eqref{ucond2} are satisfied, and the unitary equivalence between $\tilde{J}_{R}$ and $\tilde{J}$ is confirmed. Let us comment that, if $\mathbb{N}_{1}$ had a finite number of elements, then the particle-antiparticle convention corresponding to the complex structure $\tilde{J}$ would be, essentially, the opposite to that defined by our reference structure $J_{R}$, as was the case for the complex structures investigated in Ref. \cite{uf}.

Therefore, given a Fock representation with invariant complex structure $\tilde J$ that admits a nontrivial unitary implementability of the dynamics, it is either unitarily equivalent to the quantization defined by our reference complex structure $J_R$, or to an alternative one defined by $\tilde J_R$, obtained from $J_R$ by changing the convention in the notion of particles and antiparticles for an infinite number of modes. In this regard, and in the light of our discussion above, we emphasize that this change in the convention is not necessary if the attention is restricted to complex structures for which the discrepancy about the role of particle and antiparticle with respect to $J_{R}$ occurs only for a finite subset of modes. In this sense, one could argue that a further characterization of the possible invariant Fock representations for a fermion field consists in distinguishing among the distinct conventions that differ in the assignation of what is a particle and what is an antiparticle for an infinite collection of modes. It is obvious that, without additional requirements, not all the complex structures that correspond to the same convention lead to quantum theories that are unitarily equivalent, but in fact there is an infinite number of inequivalent families among them. Even when restricting all the attention to invariant Fock representations with a given convention, the problem of picking out a preferred quantization is still present. What we have shown is that, in these circumstances, it suffices to impose the extra requirement of the existence of a nontrivial unitarily implementable dynamics to indeed select a unique representation, up to unitary equivalence.

In summary, we have proven that all the invariant complex structures that allow for a nontrivial unitary implementation of the quantum dynamics in the Fock space that they define, are unitarily equivalent, up to changes in the convention for the notion of particles and antiparticles that affect an infinite collection of modes. Our criteria also select a unique field parametrization, understood as the necessary presence of a time-dependent and mode-independent scaling at dominant asymptotic order, displayed in Eq. \eqref{unith}.

\section{Conclusions}
\label{sec:Conclu}

In this work we have provided a complete proof of the uniqueness of the Fock quantization of a massive Dirac field propagating in a homogeneous and isotropic cosmological background, with spatial sections that are topologically three-spheres, under the criteria of vacuum invariance under the group of spatial isometries and unitary implementability of the dynamics. For this result, the conditions imposed on the time dependence of the background scale factor are very mild: Its third derivative must exist and must be integrable in every closed time interval of the evolution. Our work extends the results of our previous reference \cite{uf} in two directions. First of all, we have reviewed the relation between the irreducible representations of the isometry group $\text{Spin}(4)$ and the eigenspaces of the Dirac operator on the three-sphere, and specified the form that a complex structure has to adopt in order to be invariant under the action of such an isometry group. Considering only these invariant complex structures, we have then proven that the equivalence class constructed in Ref. \cite{uf} is the only one selected by the extra criterion of admitting a nontrivial implementation of the dynamics as a unitary operator on the Fock space, once a concrete convention for the notion of particles and antiparticles has been settled. In other words, having fixed such a convention, if an invariant Fock representation is unitarily inequivalent to a representative of the mentioned class, then it does not lead to a quantization where the dynamical evolution is nontrivial and can be implemented by a unitary operator on the Fock space.

As we have said, the unitary equivalence within this class is found only after adopting a convention of what is a particle and what is an antiparticle, except possibly for a finite number of modes. The need for this additional fixation can be regarded as a manifestation of the fact that two theories with a contradictory notion for particles and antiparticles, affecting an infinite number of degrees of freedom, will differ in the physical interpretation of the results if their conventions are not reconciled.

The uniqueness proof presented in this work strongly relies on the asymptotic behavior of the solutions of the Dirac equations, in the regime of large absolute value of the eigenvalues of the Dirac operator \cite{uf}. By analyzing all possible scenarios for an invariant complex structure, we have seen that the nontrivial unitary implementability of the fermion dynamics requires a very specific asymptotic behavior for the annihilation and creationlike variables. Besides, it requires also a very specific description of the fermion modes in terms of which the Dirac field decomposes, via a mode-independent and time-dependent scaling of the dominant term in the asymptotic limit of large eigenvalues, that involves the scale factor of the cosmology under study. The leading order in the mentioned asymptotics defines our reference complex structure, as given in Eq. \eqref{reference}. This asymptotic behavior of the invariant complex structures, and the associated parametrization of the Dirac field, are in fact needed for the unitary dynamics, provided that the phases of each of the fermion modes are not specifically modified (through a mode and time-dependent contribution) so that their dominant dynamical component is trivialized.

In consequence, our criteria not only fix the Fock representation in a unique manner once a suitable set of variables has been chosen to describe the field, but they actually remove as well the ambiguity in that choice arising from time-dependent linear reparametrizations. Any other time-dependent description of the fields, different at leading order from the one displayed by our reference quantization, necessarily prevents the implementability of the dynamics in a unitary way. This affects both the global scaling introduced in the decomposition \eqref{harm1} and \eqref{harm2}, and the scalings in the particle and antiparticle contributions that are induced by the time dependence of the functions $f^n_l$ and $g^n_l$ in Eq. \eqref{reference}. Therefore, the fermion field presents specific and different time-dependent scalings in its particle and antiparticle parts, scalings which are different as well for each of the two chiralities \cite{uf}. This introduces a novelty with respect to the scalar field analyses \cite{compact,torus}. In those cases, the unitarity of a nontrivial dynamics imposes a global scaling of the original field variable, such that the scaled field in practice behaves similarly to a conformally coupled field. Moreover, in the ultraviolet regime of large eigenvalues $\omega_n$, the mass contribution to the dynamics becomes negligible, so in that regime the evolution of the field is conformally invariant, owing also to the fact that the background is conformally flat. One might then wrongly believe that the unitarity of the dynamics is possible just because of this conformal invariance in the ultraviolet regime. Nevertheless, the present fermionic case shows that such belief is false. Indeed, in this case the mass term couples the two chiralities of the Dirac field, a fact that translates into a non-negligible contribution in the ultraviolet regime of the dynamics. Therefore, the evolution cannot be treated as asymptotically conformally invariant, but it is yet implementable as a unitary quantum endomorphism. The unitary evolution is possible only if one introduces the commented description of the Dirac field through a partial and time-dependent scaling at dominant order.

We have particularized our study to cosmologies with compact spatial sections that have the topology of a three-sphere. A similar analysis might be carried out in the case of other topologies. For instance, for a flat FRW model, compact sections are three-tori, for which the isometry group is the free Abelian group obtained by composing translations in each of the independent, periodic spatial directions. In this case, the Dirac operator can be specified by choosing the trivial spin structure on the torus, that is, the structure that naturally arises e.g. from parallelizing the torus with a diagonal triad \cite{SGeom,Dtorus}. Its eigenspaces provide representations of the isometry group, which can be understood as $U(1)\times U(1)\times U(1)$ acting on the space of two-component spinors. The main novelty in the analysis of this flat model is that now the irreducible representations of the isometry group are generally not in one-to-one correspondence with the eigenspaces of the Dirac operator. This is due to the fact that different inequivalent irreducible representations of the group can be realized on the same eigenspace of the Dirac operator owing to accidental degeneracy. Nonetheless, this peculiarity does not pose an obstruction to characterize the \emph{invariant} complex structures. In fact, a similar situation was dealt with satisfactorily in Ref. \cite{torus} for the case of a scalar field in a flat FRW model. Taking into account the plane wave nature of the eigenspinors of the Dirac operator on the three-torus, we expect that the study of the Dirac field can be completed along the same basic lines. Therefore, in this flat case we also expect uniqueness of the quantization (up to unitary equivalence) if we impose the condition of a unitary implementation of the dynamics, restricting the attention to invariant complex structures. Moreover, we do not anticipate serious difficulties in the generalization of the results to compact hyperbolic spatial sections, provided a detailed spectral analysis of the corresponding Dirac operator is available \cite{hyperbolic}. Let us finally comment that the direct extension of our conclusions to noncompact (flat) universes may be complicated by the typical infrared divergences associated with pair production in an infinite volume. However, note that, in the context of cosmology, we can always choose compactification scales larger than a cosmological one related to the Hubble radius. Since relevant physical results should not be sensitive to such scales, the hypothesis of compactness should not be crucial.

Beyond their interest from the point of view of QFT in curved backgrounds, the uniqueness results demonstrated in this work can prove to be useful when applied to more general physical systems. Indeed, when approaching the deep Planck regime of a cosmological scenario such as the one contemplated here, quantum gravity phenomena that exceed the context of QFT are commonly expected to arise. In such situations, e.g. in the early Universe, a plausible option to describe the physics is to treat the zero modes of the geometry (namely, in this case the homogeneous scale factor) with quantum gravity techniques, whereas the possible inhomogeneous fields are treated within a QFT approach. This was actually done in Ref. \cite{H-D} for fermion perturbations in the Planck regime of the early Universe, employing a Schr\"odinger representation for the zero modes of the geometry. Recently, this strategy has also been put forward in the context of loop quantum cosmology \cite{lqc}, by means of the so-called hybrid quantization \cite{hybr}, in order to describe scalar and tensor perturbations in the Planck regime of the early Universe \cite{hyb-pert,dress}. If a massive Dirac field is to be put in this equation, then our results provide a very specific and physically meaningful choice for the description of the field, together with its Fock quantization. This choice is relevant inasmuch as the background geometry is being quantized at the same time as the field, so that a different time-dependent parametrization of the field may result in a different prescription for the quantization of the whole system.
Another area of specialization in which possible extensions of the present work might find application is condensed matter physics. Indeed, as commented in the Introduction, it has been shown, both theoretically and experimentally, that low energy excitations in graphene obey the dynamics of a massless Dirac fermion, in $2+1$ dimensions \cite{graphene}. It would then be interesting to see whether a similar criterion to that presented here could be found for a massless Dirac particle in such a type of scenario.

\acknowledgments
This work was partially supported by the research grants MINECO Grant No. FIS2014-54800-C2-2-P from Spain, DGAPA-UNAM IN113115 and CONACyT Grant No. 237351 from Mexico, and COST Action MP1405 QSPACE, supported by COST (European Cooperation in Science and Technology). In addition, M. M.-B. acknowledges financial support from the Netherlands Organization for Scientific Research (NWO) (Grant No. 62001772). The authors thank W.D. van Suijlekom for pointing out Ref. \cite{Yasushi}.

\appendix

\section{Spinor conventions}
\label{appendixA}

We summarize here the main conventions followed in this work for the treatment of two-component spinors. First of all, the change between spinors (as well as their complex conjugates) and their algebraic duals is mediated by the antisymmetric objects $\epsilon^{AB}$, $\epsilon_{AB}$, $\epsilon^{A'B'}$, and $\epsilon_{A'B'}$, each of which is given by the matrix
\begin{align}
\begin{pmatrix}
0 & 1 \\ -1 & 0
\end{pmatrix}.\nonumber
\end{align}
In practice, these objects raise and lower spinor indices. For example, $\phi_{A}=\phi^{B}\epsilon_{BA}$ and ${\bar{\chi}}^{A'}=\epsilon^{A'B'}{\bar{\chi}}_{B'}$. Besides, tetrad vectors are related to spinors via the soldering forms $\sigma^{AA'}_{a}$, given by
\begin{align}
\sigma_{0}=-\frac{1}{\sqrt{2}}I, \qquad \sigma_{\alpha}=\frac{1}{\sqrt{2}}\Sigma_{\alpha},
\end{align}
where $I$ is the $2\times 2$ identity matrix, and $\Sigma_{\alpha}$ (with $\alpha=1,2,3$) are the three standard Pauli matrices. The spinor version of the tetrad is simply
\begin{align}
e^{AA'}_{\mu}=e^{a}_{\mu}\sigma_{a}^{AA'},
\end{align}
and then the spinor version of any spacetime tensor $T^{\mu_{1}...\mu_{n}}$ is constructed as
\begin{align}
T^{AA'...DD'}=e_{\mu_{1}}^{AA'}...e_{\mu_{n}}^{DD'}T^{\mu_{1}...\mu_{n}}.
\end{align}

Let us also introduce in this appendix the constant coefficients $\breve{\alpha}_{n}^{pq}$ and $\breve{\beta}_{n}^{pq}$ that are included in the expansions \eqref{harm1} and \eqref{harm2} of the fermion fields, in order to avoid couplings in the Einstein-Dirac action between different values of $p$. They can be regarded as the coefficients of two real matrices $\breve{\alpha}_{n}$ and $\breve{\beta}_{n}$, each of dimension $g_n=(n+1)(n+2)$. These matrices are block diagonal, with blocks given by
\begin{align}
\begin{pmatrix}
1 & 1 \\ 1 & -1
\end{pmatrix}\qquad \text{and} \qquad \begin{pmatrix}
1 & -1 \\ -1 & -1
\end{pmatrix}\nonumber
\end{align}
for $\breve{\alpha}_{n}$ and $\breve{\beta}_{n}$, respectively.

\end{document}